\documentclass[useAMS,usegraphicx,usenatbib,referee]{mn2e}

\title[Escape flux from modified shocks]
{On the escape of particles from cosmic ray modified shocks}
\author[D. Caprioli, P. Blasi and E. Amato]{D. Caprioli$^1$
\thanks{E-mail: caprioli@sns.it}, P. Blasi$^{2,3}$\thanks{E-mail:
  blasi@arcetri.astro.it} and E. Amato$^{2}$\thanks{E-mail:
  amato@arcetri.astro.it}\\ 
$^1$ SNS, Scuola Normale Superiore, Pisa\\
$^{2}$INAF-Osservatorio Astrofisico di Arcetri, 
Largo E. Fermi, 5, 50125, Firenze, Italy\\
$^{3}$Fermilab, Center for Particle Astrophysics, Batavia, IL, USA}

\begin{document}


\maketitle

\label{firstpage}

\begin{abstract}
Stationary solutions to the problem of particle acceleration at shock
waves in the non-linear regime, when the dynamical reaction of 
the accelerated particles on the shock cannot be neglected, are known 
to show a prominent energy flux escaping from the shock towards upstream 
infinity. On physical grounds, the escape of particles from the 
upstream region of a shock has to be expected in all those situations 
in which the maximum momentum of accelerated particles, $p_{max}$,
decreases with time, as is the case for the Sedov-Taylor  
phase of expansion of a shell Supernova Remnant, when both the shock 
velocity and the cosmic ray induced magnetization decrease. In this
situation, at each time $t$, particles with momenta larger than
$p_{max}(t)$ leave the system from upstream, carrying away a large
fraction of the energy if the shock is strongly modified by the
presence of cosmic rays. This phenomenon is of crucial importance
for explaining the cosmic ray spectrum detected at Earth.
In this paper we discuss how this escape flux appears in the different
approaches to non-linear diffusive shock acceleration, and especially
in the quasi-stationary semi-analytical kinetic ones.
We apply our calculations to the Sedov-Taylor phase of a typical
supernova remnant, including in a self-consistent way
particle acceleration, magnetic field amplification and the 
dynamical reaction on the shock structure of both particles and
fields. Within this framework we calculate the temporal evolution of
the maximum energy reached by the accelerated particles and of the
escape flux towards upstream infinity. The latter quantity is directly 
related to the cosmic ray spectrum detected at Earth. 
\end{abstract}

\begin{keywords}
acceleration of particles - shock waves
\end{keywords}

\section{Introduction}
Supernova Remnants (SNRs) have long been suspected to be the main sources
of galactic cosmic rays with energies up to the so called ``knee''
($E\sim 3 \times 10^{15}$ eV). The existence of a flux of energetic
particles escaping the accelerator from the upstream region is essential
if this paradigm is indeed realized in Nature. The commonly accepted view
of a typical SNR evolution is as follows. During the initial stage
after the supernova explosion, the ejecta are in free-expansion with a
high velocity $\sim 10^4$ km/s. Non-linear acceleration is expected to
be at work: efficient particle acceleration is associated with shock
modification and magnetic field amplification in a complex chain of
causality. During this stage, the maximum momentum achieved by the
accelerated particles increases with time up to $p_{max}\sim 10^{16}$ eV/c 
(see e.g. \cite{bac07} and references therein). When the mass of
the swept-up material becomes comparable with the mass of the ejecta,
the Sedov-Taylor phase begins: the shock slows down and magnetic field 
damping is expected to become more effective
(\cite{ptuzira05}). As a consequence, the maximum momentum of particles 
that can return to the shock is expected to
decrease. At each time $t$, particles with momentum exceeding the current 
$p_{max}(t)$ do not make it back to the shock and leave the system
from upstream: this is the so called ``escape flux''.   

The escape flux plays an essential role
in the formation of the cosmic ray spectrum detected at Earth. If 
there were no escape from upstream during the Sedov-Taylor phase, 
all particles accelerated in a SNR would be advected downstream and
undergo severe adiabatic energy losses before being injected into the
ISM: in this case SNRs could not be responsible for accelerating
cosmic rays with energies around the knee (\cite{crspec}, \emph{in
  preparation}, and \cite{ptuzira05}). 

While being of crucial importance during the Sedov-Taylor phase, 
it is clear that escape cannot occur during the free expansion
phase of a SNR: the only particles that can leave the system from
upstream are the ones at the highest energies, but, since $p_{max}$ is
now increasing with time, no appreciable energy flux can escape 
the accelerator during this phase.

Kinetic approaches to non-linear particle acceleration
(\cite{malkov97,maldrury,blasi1,blasi2,ab05}) allow us to calculate the
spectrum and the spatial distribution (including the absolute normalization) 
of the particles accelerated at the shock, even in the case when the
diffusion coefficient is the result of magnetic field amplification by
streaming instability induced by the accelerated particles themselves 
(\cite{ab06}).  
These approaches, as well as others that have appeared in the literature
(for instance \cite{simple}), are based on the assumption that the
acceleration process (and the shock modification) may be assumed to
reach some sort of stationarity. In all these cases the calculations show
that the shock is strongly modified by the presence of cosmic rays,
and that the spectra are concave, with a slope at momenta close to
the maximum momentum $p_{max}$ which is flatter than $p^{-4}$. All these
calculations, independently of the techniques used to solve the equations, 
predict an escape flux of particles (and energy) towards upstream infinity: 
the shock becomes radiative, which is one of the very reasons why the shock 
modification becomes effective (namely the total compression factor becomes 
larger than 4).  

It is worth recalling that the assumption of stationarity was widely 
used also in the framework of two-fluid models
(\cite{drury81p,drury81}), and its relevance in the case of modified
shocks, leading to particle spectra harder than $p^{-4}$ at high
energies, was discussed by \cite{mv96} in the context of the
  so-called renormalized two-fluid models.
It should be stressed, indeed, that the escape flux can arise both 
in the linear and non-linear regime of particle acceleration. The important
difference between the two regimes, however, is that within the former 
the escaping particles carry away a negligible energy flux, while for
modified shocks the bulk of the energy in the form of cosmic rays is 
stored in the highest energy particles and leaves the system with them. 
 
The main weak point of the assumption of quasi-stationarity is that it
leads to artificial escape fluxes in situations, such as the free
expansion phase of a SNR, where escape would be unphysical (at least
in the idealized cases of plane and perfectly spherical shocks). 
In this case, the
appearance of an escape flux signals for the need for a fully
time dependent approach. In other words, what appears as an escape
flux is the energy which is channeled into particles with ever
increasing momentum. During the Sedov-Taylor phase, as emphasized
earlier, escape is in fact possible if the magnetic field in the shock
region decreases. In this case, the leakage of particles towards
upstream allows the system to relax to a quasi-stationary situation. 
 
A number of time-dependent studies of non-linear shock dynamics 
have been presented in the literature (e.g. \cite{fg87}), in
particular for SNRs (e.g. \cite{bv97,kj06} and references
therein). All these calculations aim at solving the transport
equations and the equation of fluid dynamics numerically. Most of
these papers have however concentrated on the spectrum of accelerated
particles rather than on the escape flux. In fact, in many of these
papers the magnetic field is introduced by hand and is not a function
of time, thereby not leading to a decrease with time of the maximum
momentum of accelerated particles. In other cases the boundary
conditions are imposed at upstream infinity, therefore the escape
flux, even if it is present, would appear in the form of concentration
of high energy particles at large distances from the shock. In these
cases a recipe is needed to label these particles as {\it escaping}
the system (see for instance \cite{kamae}).

The identification of the escape flux is all but a mathematical
detail: as discussed by \cite{ptuzira05}, the escape flux is crucial for
establishing a connection between SNRs and the origin of cosmic rays,
in that it allows the escaping particles to 
avoid the adiabatic losses related to the expansion of the supernova
shell, and therefore become cosmic rays at the knee. In the absence of
escape towards upstream infinity at the beginning of the Sedov-Taylor
phase, SNRs cannot be the sources of cosmic rays up to the knee.  

The contribution of the escape flux to the galactic cosmic ray
spectrum was first estimated by \cite{ptuzira05}. These authors
included the magnetic field damping and the consequent decrease of
$p_{max}$ during the Sedov-Taylor expansion, but in their study the
shock structure and particle spectrum are fixed rather than
self-consistently calculated. 

In this paper we present the first attempt at carrying out a
calculation of the escape flux within the framework of a kinetic
approach to non-linear shock acceleration, self-consistently including
particle acceleration, magnetic field amplification and the dynamical
reaction of both on the shock. We focus our attention on shocks in
shell type SNRs, discussing the physical meaning of the escape flux
during the different phases of their evolution. In particular the flux
of energetic particles leaving the remnant during the Sedov-Taylor phase
is explicitely computed and its phenomenological implications for the 
origin of cosmic rays are discussed. 

The paper is organized as follows: in \S \ref{sec:stat} we discuss the
implications of the assumption of stationarity of the acceleration
process. In \S \ref{sec:cons} we discuss the escape flux based on the
most general version of the conservation equations. In \S
\ref{sec:escape} we discuss how the escape flux is connected to the
existence of a maximum momentum in the distribution of accelerated
particles. In \S \ref{sec:snr} we apply our calculations to the
different stages of evolution of a supernova remnant. We conclude in
\S \ref{sec:concl}.

\section{The assumption of stationarity} 
\label{sec:stat}

The standard solution of the stationary transport equation
\begin{equation}
u(x)\frac{\partial f(x,p)}{\partial x}=
\frac{\partial}{\partial x}\left[D(p)\frac{\partial f(x,p)}{\partial x}\right]
+ \frac{p}{3}\frac{du}{dx}\frac{\partial f}{\partial p} + Q  
\end{equation}
leads, in the test-particle regime, to the well known power-law spectrum 
of accelerated particles $f(p)\propto p^{-\alpha}$, with $\alpha=3r/(r-1)$ 
where $r$ is the compression factor at the shock. 

The power law extends to infinitely large momenta. Since for ordinary 
non-relativistic gaseous shocks $r<4$ (namely $\alpha>4$), the total energy in 
the form of accelerated particles remains finite. This solution is found by 
imposing as boundary condition at upstream infinity ($x=-\infty$)
that $f(-\infty)=0$ and $\partial f(-\infty)/\partial x=0$. 

If the boundary condition $f(x=x_0)=0$ is used, instead, at some finite 
distance $x_0<0$ upstream, the solution of the transport equation is easily 
calculated to be
\begin{equation}
f(x,p)=\frac{f_0(p)}{1-\exp\left(\frac{u_1 x_0}{D(p)} \right)}
\left[\exp\left(\frac{u_1 x}{D(p)}\right) -  \exp\left(\frac{u_1
x_0}{D(p)}\right) \right],
\end{equation}
where 
\begin{equation}
f_0(p)=K\exp\left\lbrace -\frac{3u_1}{u_1-u_2} \int_{p_{inj}}^{p}
\frac{dp'}{p'} \frac{1}{1-\exp\left(\frac{x_0 u_1}{D(p')} \right)} 
\right\rbrace  .
\end{equation}
In the case of Bohm diffusion, $D(p)=D_0 (p/m_p c)$ and one obtains:
\begin{equation}
f_0(p)=K\exp\left\lbrace -\frac{3u_1}{u_1-u_2} \int_{p_{inj}}^{p}
\frac{dp'}{p'} \frac{1}{1-\exp\left(-\frac{p_*}{p'} \right)} \right\rbrace  .
\end{equation}
where $p_{*}=|x_0| u_1 m_p c/D_0$. Now one can show that for $p\ll p_*$,
$f_0(p)\propto (p/p_*)^{-3r/(r-1)}$, with $r=u_1/u_2$, the standard result.
However, for $p\gg p_*$,
$f_0(p)\propto\exp\left[-\frac{3r}{r-1}\frac{p}{p_*}\right]$. 
The quantity $p_{max}=p_*(r-1)/3r$ plays the role of maximum momentum of
the accelerated particles. 

\begin{figure}
\centering
\includegraphics[width=0.9\textwidth]{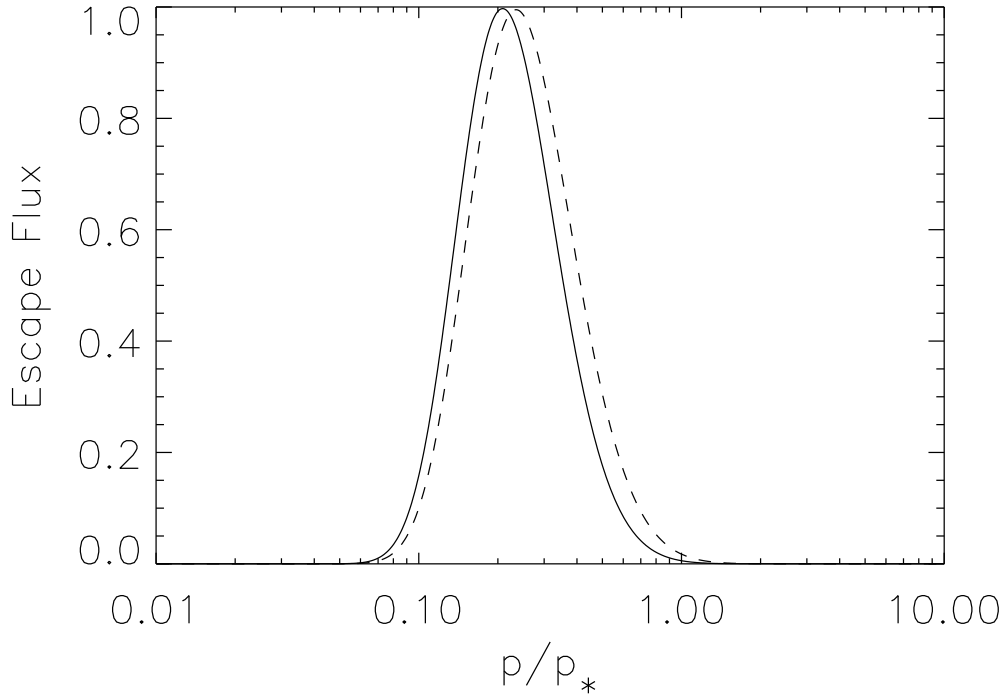}
\caption{We plot the escape flux $\phi(x_0, p)$ as a function of momentum.
The curves refer to two different values of the shock compression ratio:
$r=4$ (solid line) and $r=7$ (dashed line). The computation is carried out
in the test-particle regime. The $x$-axis is in units of the reference 
momentum $p_*=r/(r-1) p_{max}$, while units along the $y$-axis are 
arbitrary}
\label{fig:escapeflux}
\end{figure}

This simple example shows how a maximum momentum can be obtained in a
stationary approach only by imposing the boundary condition at a
finite boundary. 
Physically this corresponds to particles' escape, as
shown by the fact that the flux of particles at $x=x_0$ is 
\begin{equation}
\phi(x_0,p) = u_1 f(x_0,p) - D(p) \frac{\partial f(x_0)}{\partial x} = 
-\frac{u_1 f_0(p)}{1-\exp\left( \frac{u_1 x_0}{D(p)} \right)}
\exp\left(\frac{u_1 x_0}{D(p)} \right)<0.
\end{equation} 
The fact that $\phi(x_0,p)<0$ shows that the flux of particles is
directed towards upstream infinity. Moreover, the escape flux as a 
function of momentum, $\phi(x_0,p)$, is negligible for all $p$ with 
the exception of a narrow region around $p_{max}$: only particles 
with momentum close to $p_{max}$ can escape the system towards upstream 
infinity. The escape flux as a function of momentum is plotted in 
Fig.~\ref{fig:escapeflux} for two values of the compression factor, 
$r=4$ (solid line) and $r=7$ (dashed line). The normalizations are 
arbitrary, since the calculations are carried out in the context of 
test particle theory. The latter value of $r$ cannot be realized at
purely gaseous shocks, but we have adopted this value to mimic 
the effect of shock modification, which leads to total compression 
factors larger than 4.

The escape phenomenon is basically irrelevant in the test-particle regime,
because of the negligible fraction of energy carried by particles with 
$p\sim p_{max}$, but it becomes extremely important in the calculation 
of the shock modification induced by accelerated particles. For strongly 
modified shocks, the slope of the spectrum at high energies is flatter than 
$p^{-4}$ and the fraction of energy that leaves the system towards upstream 
infinity may dominate the energy budget. This is the escape flux which 
appears in all approaches to cosmic ray modified shocks.

In the context of kinetic calculations of the shock modification in the
stationary regime, the escape flux appears however not as a consequence of
imposing a boundary condition at a finite distance upstream, but rather as an
apparent violation of the equation of energy conservation (\cite{simple}), that
requires the introduction of an escape term at upstream infinity. 
In the next section we discuss this effect, which reveals the true nature of
the escape flux, as related to the form of the conservation equations and the
assumption of stationarity.

\section{Conservation equations and escape flux}
\label{sec:cons}

In this section we rederive the conservation equations for cosmic
ray modified shocks in their general form, in order to emphasize the
mathematical origin of the escape flux.

The time dependent conservation equations in the presence of
accelerated particles at a shock can be written in the following form:
\begin{eqnarray}
\frac{\partial \rho}{\partial t} &=& - \frac{\partial (\rho u)}{\partial x}\\ 
\frac{\partial (\rho u)}{\partial t} &=& -\frac{\partial}{\partial x}
\left[ \rho u^2 + P_g + P_c+P_W \right]\\ 
\frac{\partial}{\partial t}\left[ \frac{1}{2}\rho u^2 +
  \frac{P_g}{\gamma_g-1} \right] &=& 
-\frac{\partial}{\partial x}\left[ \frac{1}{2}\rho u^3 +
  \frac{\gamma_g P_g u}{\gamma_g-1} \right] - u \frac{\partial}{\partial
x}\left[ P_c+P_W\right]+\Gamma E_W\ .
\label{eq:conserva}
\end{eqnarray}
Here $P_g$, $P_c$ and $P_W$ are respectively the gas pressure, the cosmic 
ray pressure and the pressure in the form of waves. $E_W$ is the energy 
density in the form of waves and $\Gamma$ is the rate at which the 
background plasma is heated due to the damping of waves onto the plasma. 
The rate of change of the gas temperature is related to $\Gamma E_W$ through:
\begin{equation}
 \frac{\partial P_g}{\partial t} + u \frac{\partial P_g}{\partial x} + \gamma_g
P_g \frac{d u}{d x} = (\gamma_g - 1) \Gamma E_W.
\end{equation} 
The cosmic ray pressure can be calculated from the transport equation:
\begin{equation}
\frac{\partial f(t,x,p)}{\partial t} + \tilde u(x) \frac{\partial f(t,x,p)}{\partial x} = 
\frac{\partial}{\partial x}\left[ D(x,p) \frac{\partial f(t,x,p)}{\partial
    x}\right] + \frac{p}{3}\frac{\partial f(t,x,p)}{\partial p} \frac{d\tilde u(x)}{dx}, 
\end{equation}
where we put $\tilde u(x) = u(x) - v_W(x)$ and $v_W(x)$ is the wave velocity.
For our purposes here we are neglecting the injection term. 

Multiplying this equation by the kinetic energy $T(p)=m_p c^2
(\gamma-1)$, where $\gamma$ is the Lorentz factor of a particle
with momentum $p$, and integrating the transport equation in momentum,
one has:
\begin{equation}
\frac{\partial E_c}{\partial t} + \frac{\partial (\tilde u E_c)}{\partial x} = 
\frac{\partial}{\partial x}\left[ \bar D \frac{\partial E_c}{\partial
    x}\right] - P_c \frac{d \tilde u}{d x},
\label{eq:cr0}
\end{equation}
where
\begin{equation}
E_c=\int_0^{\infty} dp\,4\pi p^2\, T(p) f(p)\, \qquad {\rm and}\, \qquad
P_c=\int_0^{\infty} dp\, \frac{4\pi}{3} p^3 v(p) f(p)
\end{equation}
are the energy density and pressure in the form of accelerated
particles. Moreover we introduced the mean diffusion coefficient:
\begin{equation}
\bar D(x) = \frac{\int_0^{\infty}4\pi p^2 T(p) D(p) \frac{\partial
    f}{\partial x}}{\int_0^{\infty}4\pi p^2 T(p) \frac{\partial
    f}{\partial x}}
\end{equation}
The only assumption that we made here is that $f(p)\to 0$ for $p\to \infty$. 
We shall comment later (\S \ref{sec:escape}) on what would 
happen if the spectrum were truncated at some fixed $p_{max}$, instead.

Introducing the adiabatic index for cosmic rays $\gamma_c$
as $E_c=P_c/(\gamma_c-1)$, we can rewrite Eq.~\ref{eq:cr0} as
\begin{equation}
\frac{\partial E_c}{\partial t} + \frac{\partial}{\partial x}
\left[ \frac{\gamma_c \tilde u P_c}{\gamma_c-1} \right] =
\frac{\partial}{\partial x}\left[ \bar D \frac{\partial E_c}{\partial
    x}\right] + \tilde u \frac{d P_c}{d x},
\label{eq:cr}
\end{equation} 
and use it to derive $u\partial P_c/\partial x$. In this way
Eq.~\ref{eq:conserva} becomes: 
\begin{eqnarray}
\lefteqn{\frac{\partial}{\partial t}\left[ \frac{1}{2}\rho u^2 +
  \frac{P_g}{\gamma_g-1} + E_c \right] =}  \nonumber \\ & &
=-\frac{\partial}{\partial x}\left[ \frac{1}{2}\rho u^3 +
  \frac{\gamma_g P_g u}{\gamma_g-1} + \frac{\gamma_c P_c
    \tilde u}{\gamma_c-1} \right] + \frac{\partial}{\partial x}\left[ \bar
  D(x) \frac{\partial E_c}{\partial x} \right] - v_W \frac{\partial
P_c}{\partial x} - u \frac{\partial P_W}{\partial x} + \Gamma E_W.
\label{eq:compl}
\end{eqnarray}

At this point we can make use of the equation describing the evolution of the
waves:
\begin{equation}
\frac{\partial E_W}{\partial t} + \frac{\partial F_W}{\partial x} = u
\frac{\partial P_W}{\partial x} + \sigma E_W - \Gamma E_W,
\end{equation} 
where $\sigma$ is the growth rate of waves, integrated over wavenumber 
and $F_W$ is the flux of energy in the shape of waves. 
These quantities can be calculated once it is known how particles with 
given momentum $p$ interact with waves with wavenumber $k$ and how the turbulence energy is shared between the kinetic and the
properly magnetic contributions. 
Substituting into Eq.~\ref{eq:compl} we get:
\begin{eqnarray}
\lefteqn{\frac{\partial}{\partial t}\left[ \frac{1}{2}\rho u^2 +
  \frac{P_g}{\gamma_g-1} + E_c +E_W \right] = } \nonumber \\ & &
=-\frac{\partial}{\partial x}\left[ \frac{1}{2}\rho u^3 +
  \frac{\gamma_g P_g u}{\gamma_g-1} + \frac{\gamma_c P_c
    \tilde u}{\gamma_c-1} +F_W\right] + \frac{\partial}{\partial x}\left[ \bar
  D(x) \frac{\partial E_c}{\partial x} \right] - v_W \frac{\partial
P_c}{\partial x} + \sigma E_W.
\label{eq:compl1} 
\end{eqnarray}
In the case of Alfv\'en waves resonant with the Larmor radius of the
accelerated particles, one has $v_W=v_A=B/(4\pi\rho)^{1/2}$ and
(\cite{skilling75c}):
\begin{equation}
 \sigma E_W = v_A \frac{\partial P_c}{\partial x},
\label{eq:sig}
\end{equation} 
so that the energy conservation equation reads
\begin{eqnarray}  
\lefteqn{\frac{\partial}{\partial t}\left[ \frac{1}{2}\rho u^2 +
  \frac{P_g}{\gamma_g-1} + E_c +E_W \right] =}  \nonumber \\ & &
=-\frac{\partial}{\partial x}\left[ \frac{1}{2}\rho u^3 +
  \frac{\gamma_g P_g u}{\gamma_g-1} + \frac{\gamma_c P_c
    \tilde u}{\gamma_c-1} +F_W\right] + \frac{\partial}{\partial x}\left[ \bar
  D(x) \frac{\partial E_c}{\partial x} \right].
\label{eq:compl2} 
\end{eqnarray}

In the general case of waves other than resonant Alfv\'en waves,
Eq.~\ref{eq:sig} does not hold and even the equality $v_W=v_A$ 
may be questionable: as a consequence one cannot use the standard 
Eq.~\ref{eq:compl2}, while Eq.~\ref{eq:compl1} is still correct.
However, in order to be able to solve the problem, the detailed form
of the cosmic ray transport equation, as well as an expression
analogous to Eq.~\ref{eq:sig}, relating the growth of the wave energy
to the cosmic ray dynamics, are still needed.
It is worth stressing, in fact, that the magnetic turbulence has often
been proposed to not show up in the form of standard resonant Alfv\'en
waves, but rather as non-resonant purely growing modes \cite{bell04},
or as generic ``magnetic structures'' in the phenomenological model of
\cite{bell-lucek01}.  
This also causes the connection between $F_W$ and $P_W$ to be generally 
different from the standard $F_W\approx 3uP_W$, thus the contribution
of these waves to the energy conservation equation does not necessarily 
lead to Eq.~\ref{eq:compl2}, which was nevertheless used by \cite{vladi06} 
in the implementation of the scenario suggested by \cite{bell-lucek01}.

In the following we limit ourselves to the case of Alfv\'en waves, which 
interact resonantly with particles, since in this case the calculations 
are all well defined. 

Notice that in the stationary regime, Eq.~\ref{eq:cr}, integrated around the
subshock leads to
\begin{equation}
\frac{\gamma_c}{\gamma_c - 1} \tilde u P_c - \bar D \frac{d E_c}{dx} = \rm
constant,
\end{equation}
because of the continuity of the cosmic ray distribution function. On the
other hand, Eq.~\ref{eq:compl2} (again in the stationary case) , once 
integrated around the shock, leads to conclude that:
\begin{equation}
\frac{1}{2}\rho u^3 + \frac{\gamma_g P_g u}{\gamma_g-1} + F_W = \rm constant.
\end{equation}
In other words, at the subshock the energy fluxes of the gaseous and
cosmic ray components are conserved separately. This is what is
usually meant when we refer to the subshock as an ordinary gas shock.
In the following we use the stationary version of Eq.~\ref{eq:compl2}:

\begin{equation}
\frac{\partial}{\partial x}\left[ \frac{1}{2}\rho u^3 + \frac{\gamma_g P_g
u}{\gamma_g-1} + \frac{\gamma_c P_c \tilde u}{\gamma_c-1} +F_W - \bar D(x)
\frac{\partial E_c}{\partial x}   \right] = 0.
\label{eq:stat} 
\end{equation} 

\section{Escape flux and the need for a $p_{max}$}
\label{sec:escape}

In non-linear theories of particle acceleration the need for a maximum
momentum is dictated by the fact that the spectrum at large momenta
becomes harder than $p^{-4}$, so that in the absence of a high $p$
cutoff the energy content of the accelerated particle distribution 
would diverge. Before this happens the dynamical reaction of the 
accelerated particles would inhibit further
acceleration. In most approaches to non-linear calculations
(\cite{malkov97,malkov00,blasi1,blasi2,simple}), the maximum
momentum is a given parameter, taken together with the assumption of
stationarity of the acceleration process. The transport equation is
then solved between the shock and upstream infinity. Both in the
downstream region and at upstream infinity one has $D\partial
f/\partial x=0$. Moreover, at upstream infinity there are no
accelerated particles ($P_c=0$) so that Eq.~\ref{eq:stat} becomes: 
\begin{equation}
\frac{1}{2}\rho_2 u_2^3 +
\frac{\gamma_g P_{g,2} u_2}{\gamma_g-1} + \frac{\gamma_c P_c u_2}{\gamma_c-1}
+ F_W = \frac{1}{2}\rho_0 u_0^3 + \frac{\gamma_g P_{g,0} u_0}{\gamma_g-1}
\label{eq:stat1}
\end{equation}
None of the calculations of particle acceleration at modified shocks 
carried out so far satisfies Eq.~\ref{eq:stat1} unless it is 
{\it completed} with an escape flux $F_{esc}$ such that
\begin{equation}
\frac{1}{2}\rho_2 u_2^3 +
  \frac{\gamma_g P_{g,2} u_2}{\gamma_g-1} + 
\frac{\gamma_c P_c u_2}{\gamma_c-1} + F_W = 
\frac{1}{2}\rho_0 u_0^3 +
  \frac{\gamma_g P_{g,0} u_0}{\gamma_g-1} - F_{esc}.
 \label{eq:stat2}
\end{equation}
Unfortunately, as showed above, this apparently harmless step is
inconsistent with $f(p,x)$ being a solution of the time-independent
transport equation. In fact this is not surprising since, as we
stressed above, the solution of such equation cannot be characterized
by a finite $p_{max}$ when the boundary condition of vanishing $f(p,x)$ 
and $\partial f/\partial x$ is imposed at upstream infinity. 

On the other hand, as also discussed by \cite{mv96} in the
context of the so-called renormalized two-fluid models, the
requirement that there is a maximum momentum leads to the appearance
of an additional term on the right hand side of Eq. \ref{eq:cr0},
which is  
\begin{equation}
+\frac{1}{3} \frac{d\tilde u}{dx} \left[ 4\pi p^3 T(p) f(p,x)
  \right]_{p=0}^{p=p_{max}}. 
\end{equation}
This reflects in an additional term in Eq. \ref{eq:stat2}:
\begin{eqnarray}
\lefteqn{
\int_{-\infty}^{0^+} dx \frac{1}{3}\frac{du}{dx} 4\pi p_{max}^3 T(p_{max})
  f(p_{max},x) =} \\ & &
   \int_{-\infty}^{0^-} dx \frac{1}{3}\frac{du}{dx} 4\pi
  p_{max}^3 T(p_{max}) f(p_{max},x) + (u_2-u_1)\frac{4\pi}{3}
  p_{max}^3 T(p_{max}) f_0(p_{max}).\nonumber
\end{eqnarray}
Since $du/dx<0$ in the precursor, this term is negative and
numerically coincides with the escape flux $F_{esc}$.

As an alternative to obtaining the escape flux in this way, one could
derive it by assuming that there is no $p_{max}$ imposed by hand, and
that a maximum momentum results from imposing the boundary condition
at a finite distance upstream, rather than at upstream infinity
(e.g. \cite{vladi06}).  

In this second case $D\partial f/\partial x$ does not vanish at the
upstream boundary and an escape flux appears in a natural way, namely
\begin{equation}
\phi_{esc} = u_0 f(x_0,p) - D \left[\frac{\partial f}{\partial
    x}\right]_{x=x_0} = - D \left[\frac{\partial f}{\partial
    x}\right]_{x=x_0} < 0,
\end{equation}
and the energy escape flux $F_{esc}$ is related to $\phi_{esc}$ through
\begin{equation}
F_{esc}=\int_{p_{inj}}^{p_{max}}4 \pi\ p^2\ dp\ \phi_{esc}(p)\ T(p)\ .
\end{equation}

In other words, no artificial escape flux needs to be introduced if a
boundary condition is imposed at a finite distance upstream, or if, as
an alternative, the fully time dependent solution of the problem can
be found.

In next section we explore the consequences of the existence of an
escape flux for the origin of cosmic rays in supernova remnants. 

\section{Physical meaning of the escape flux for supernova remnants}
\label{sec:snr}

The acceleration process in supernova remnants is expected to work 
in qualitatively different ways during the free expansion and the
Sedov-Taylor phases. Here we restrict our attention to the propagation 
in a spatially uniform interstellar medium. During the free expansion 
phase the velocity of the shell remains constant and the maximum 
momentum grows in time in a way that depends on the growth of the 
turbulent magnetic field in the upstream region. During this phase
particles cannot escape. Nevertheless the 
standard approaches to the calculation of the shock modification would 
lead to predict an escape flux, a symptom of the need to carry out fully 
time dependent calculations to treat this expansion regime. The lack 
of particles' escape implies an increase in the maximum momentum of 
the accelerated particles. This trend ends at the beginning of the 
Sedov-Taylor phase, when the inertia of the swept up material slows 
down the expanding shell. Physically, this is the reason why we expect 
that the highest energies for particles accelerated in SNRs are reached 
at the beginning of the Sedov-Taylor phase.

During this phase, the shock velocity decreases and the magnetic
field amplification upstream, as due to streaming instability, becomes 
less efficient. The generation of magnetic turbulence via streaming 
instabilities may proceed through either resonant (\cite{bell78a,bell78b}) 
or non-resonant (\cite{bell04}) coupling between particles and waves and 
the two channels are likely to dominate at different times in the history 
of the supernova remnant (\cite{pelletier06,ab08}), as discussed below.

This general picture leads to a maximum momentum that decreases with time
and to particles' escape towards upstream infinity: particles of momentum
$p_{max}(t_1)$ do not make it back to the shock at a time $t_2>t_1$. In 
other words, during the time interval between $t_1$ and $t_2$, particles 
with momentum between $p_{max}(t_1)$ and $p_{max}(t_2)$ escape from the 
system. This happens at any time, and a net flux of particles (and
energy) towards upstream infinity is realized. At any given time $t$ the
spectrum of particles that escape is highly peaked around $p_{max}(t)$ 
(see Fig.~\ref{fig:escapeflux} for the test-particle case). 
The spectrum of accelerated particles that is confined in the accelerator and
advected towards downstream is cut off at a gradually lower maximum momentum,
and this should reflect in the spectrum of secondary radiation, especially
gamma-rays. The particles trapped downstream will also eventually escape the
system, but at the time this happens they will have been affected by adiabatic
losses due to the expansion of the shell. Therefore this part of the escaping
flux, which will reflect the history of the remnant, is suspected to 
play a particularly important role only at the lowest energies in the cosmic
ray spectrum at earth. 

The flux of high energy cosmic rays, close to the knee region, as we
see below, is mainly generated during the early Sedov-Taylor
phase and is made of particles 
escaping the accelerators from upstream. The actual flux of diffuse  
cosmic rays observed at the Earth results from the integration over
time of all the instantaneous spectra of escaping particles, each peaked at
$p_{max}(t)$, 
and from the superposition of the supernova explosions that could
contribute. This integration is affected by the accelerator
properties, by the dynamics of the expanding shell and by the damping
processes that may affect the way the magnetic field is amplified by
streaming instabilities at any given time  (\cite{ptuzira05}). 

There is also another implication of the line of thought illustrated
above: the spectrum of particles that escape, as integrated over time
during the Sedov-Taylor phase, does not need to be identical to the
spectrum of particles advected towards downstream. However, the latter
are the particles which are responsible for the production of
secondary radiation (radio, X-rays, gamma rays): the concave spectra
predicted by the non-linear theory of particle  acceleration and to
some extent required to explain observations, might not be reflected
in a concavity of the spectrum of escaping particles.  

An estimate of the scalings of the relevant quantities during the ST
phase can be found as follows. The radius and velocity of the
expanding shell can be written as:
\begin{equation}
R_{sh}(t) = 2.7\times 10^{19} {\rm cm} \left(\frac{E_{51}}{n_0}
\right)^{1/5} t_{kyr}^{2/5} 
\end{equation}
\begin{equation}
V_{sh}(t) = 4.7 \times 10^{8} {\rm cm/\rm s}\
\left(\frac{E_{51}}{n_0} \right)^{1/5}
t_{kyr}^{-3/5}, 
\end{equation}
where $E_{51}$ is the kinetic energy of the shell in the free expansion 
phase in units of $10^{51}$ erg and $n_0$ is the number density of
the plasma upstream in units of ${\rm cm}^{-3}$. Here we assumed the
standard Sedov-Taylor time-scaling of $R_{sh}$ and $V_{sh}$, but the
reader should bear in mind that the adiabatic solution may be affected
by the fact  that in this phase the shock is radiating energy in the
form of cosmic rays. The maximum energy is estimated by requiring that
the diffusion length upstream equals some fraction (say 10\%) of
$R_{sh}(t)$. If the diffusion coefficient is assumed to be Bohm-like
and the magnetic field close to the shock is $\delta B(t)$, one
obtains: 
\begin{equation}
E_{max}(t) = 3.8\times 10^{4}\ \delta B_{\mu G}(t) \left(\frac{E_{51}}{n_0}
\right)^{2/5}\ t_{kyr}^{-1/5}\ GeV.   
\end{equation}
The magnetic field in the shock vicinity is amplified by streaming
instability, induced by the accelerated particles both resonantly and
non-resonantly. Let us introduce the acceleration efficiency as 
a function of time: $\xi_c(t)=P_c(t)/(\rho_0 V_{sh}(t)^2)$. In terms
of $\xi_c$, the strength of the resonantly amplified magnetic field 
at the saturation level can be estimated as: 
$\delta B^2= 8\pi\rho_0 V^2 \xi_c/M_A$
($M_A$ is the Alfv\'en Mach number), which leads to:
\begin{equation}
\delta B(t) = 65\  n_0^{1/4}\ B_{0,\mu G}^{1/2}\
\left(\frac{E_{51}}{n_0} \right)^{1/10} t_{kyr}^{-3/10}\ \xi_c(t)^{1/2}\ 
\mu G\ .
\label{eq:delB1t}
\end{equation}
In a similar way, the strength of the field in the case of non-resonant
amplification can be estimated from 
$\delta B^2=2\pi \rho_0 (V_{sh}(t)^3/c) \xi_c(t)$ and leads to:
\begin{equation}
\delta B(t) = 198\ n_0^{1/2} \left(\frac{E_{51}}{n_0} \right)^{3/10}
t_{kyr}^{-9/10}\ \xi_c(t)^{1/2}\ \mu G\ .
\label{eq:delB2t}  
\end{equation}

In general the two channels of magnetic field amplification work
together but the non-resonant channel dominates at earlier times and
leads to stronger magnetic field amplification. 

The maximum momentum in the two cases is as follows:
\begin{equation}
E_{max}(t) = 2.5 \times 10^6 \left(\frac{E_{51}}{n_0} \right)^{1/2}  n_0^{1/4}
B_{0,\mu G}^{1/2}\ \xi_c(t)^{1/2}\ t_{kyr}^{-1/2}\ GeV,
\label{eq:emax1t}
\end{equation}
in the resonant case, and 
\begin{equation}
E_{max}(t) = 7.3 \times 10^6 \left(\frac{E_{51}}{n_0} \right)^{7/10}
n_0^{1/2}\ \xi_c(t)^{1/2}\ t_{kyr}^{-11/10}\ GeV
\label{eq:emax2t}
\end{equation}
in the non-resonant regime. 

In the naive assumption that the acceleration efficiency is constant
in time, we see that $E_{max}(t)$ scales with time as $t^{-11/10}$ at
earlier times and as $t^{-1/2}$ at later times, when resonant
scattering dominates. In actuality the scalings will be more complex
because of the non-linear effects (especially the formation of a
precursor upstream) induced by accelerated particles, which also lead
to a time dependence of $\xi_c(t)$. 

As discussed in the previous sections, it is not clear how to describe the
non-resonant waves in the context of the conservation equations. A calculation
of the dynamical effect of these modes on the shock is therefore not reliable
at the present time. For this reason, here we confine ourselves to the
investigation of the effects of resonant waves, for which there is no
ambiguity. It is however worth keeping in mind that the introduction of the
non-resonant waves is likely to result in significantly higher maximum energies at the early
stages of the Sedov-Taylor phase.  

\begin{figure}
\resizebox{\hsize}{!}{
\includegraphics{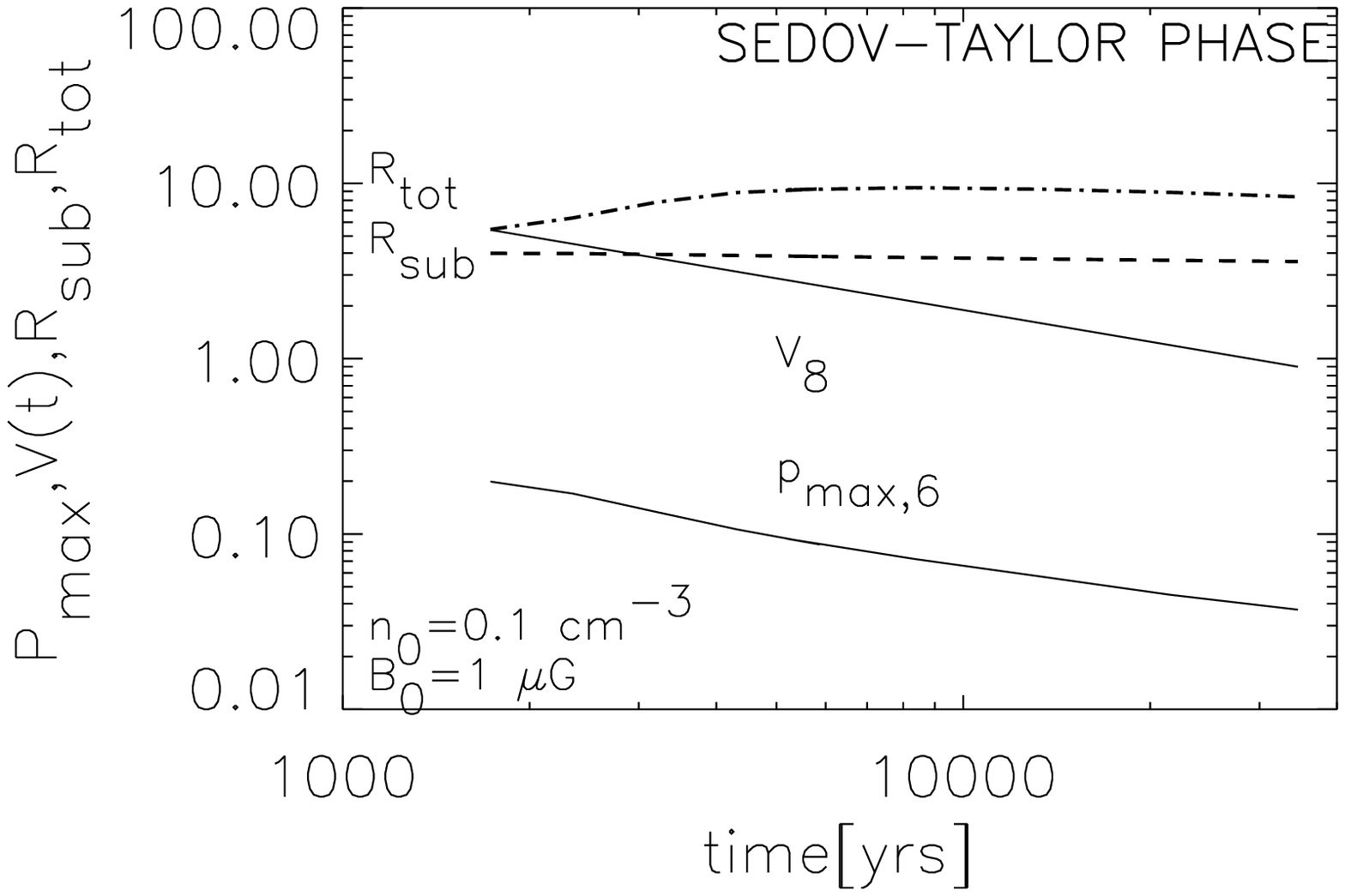}
\includegraphics{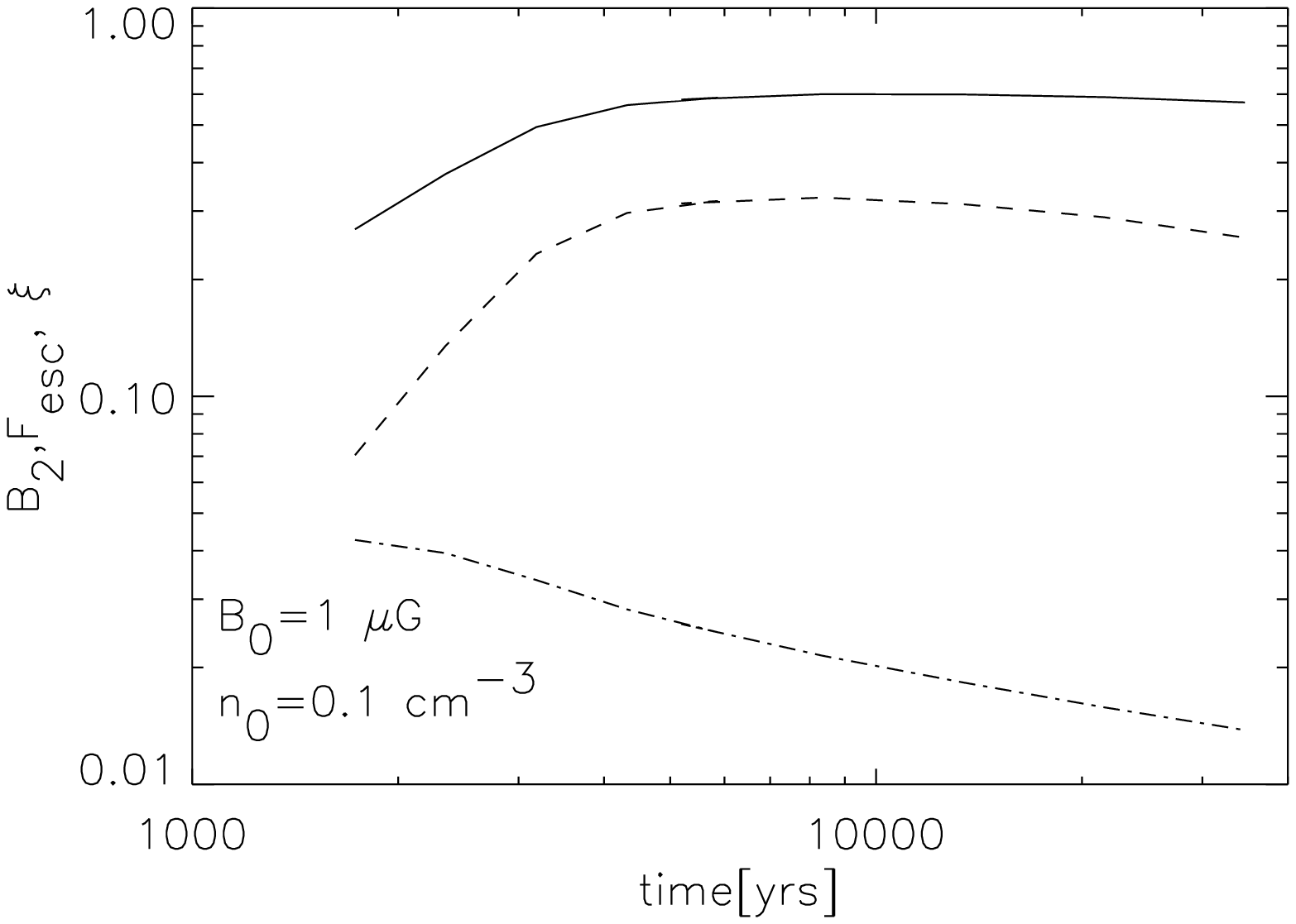}
}
\caption{Left panel: time dependence of the maximum momentum of
  accelerated particles (solid curve labeled as $P_{max,6}$) in units
  of $10^6 m_p c$, of the shock velocity in units of $10^8\ {\rm
  cm/s}$ (solid curve labeled as $V_8$), of the compression factor at
  the subshock $R_{sub}$ (dashed curve) and of the total compression
  factor $R_{tot}$ (dot-dashed curve), during the Sedov-Taylor phase. 
Right panel: time dependence of the magnetic field strength downstream
  of the subshock in units of $10^3\ \mu G$ (dot-dashed curve), of the
  escape flux normalized to $\rho_0 V_{sh}^3/2$ (dashed curve) and of
  the cosmic ray pressure at the shock location normalized to $\rho_0 V_{sh}^2$.
The magnetic field strength, $B_0$, and the number density of the background 
plasma, $n_0$, at upstream infinity are taken to be $B_0=1\ \mu G$ and 
$n_0=0.1\ {\rm cm}^{-3}$.
}
\label{fig:B1n01}
\end{figure}

\begin{figure}
\resizebox{\hsize}{!}{
\includegraphics{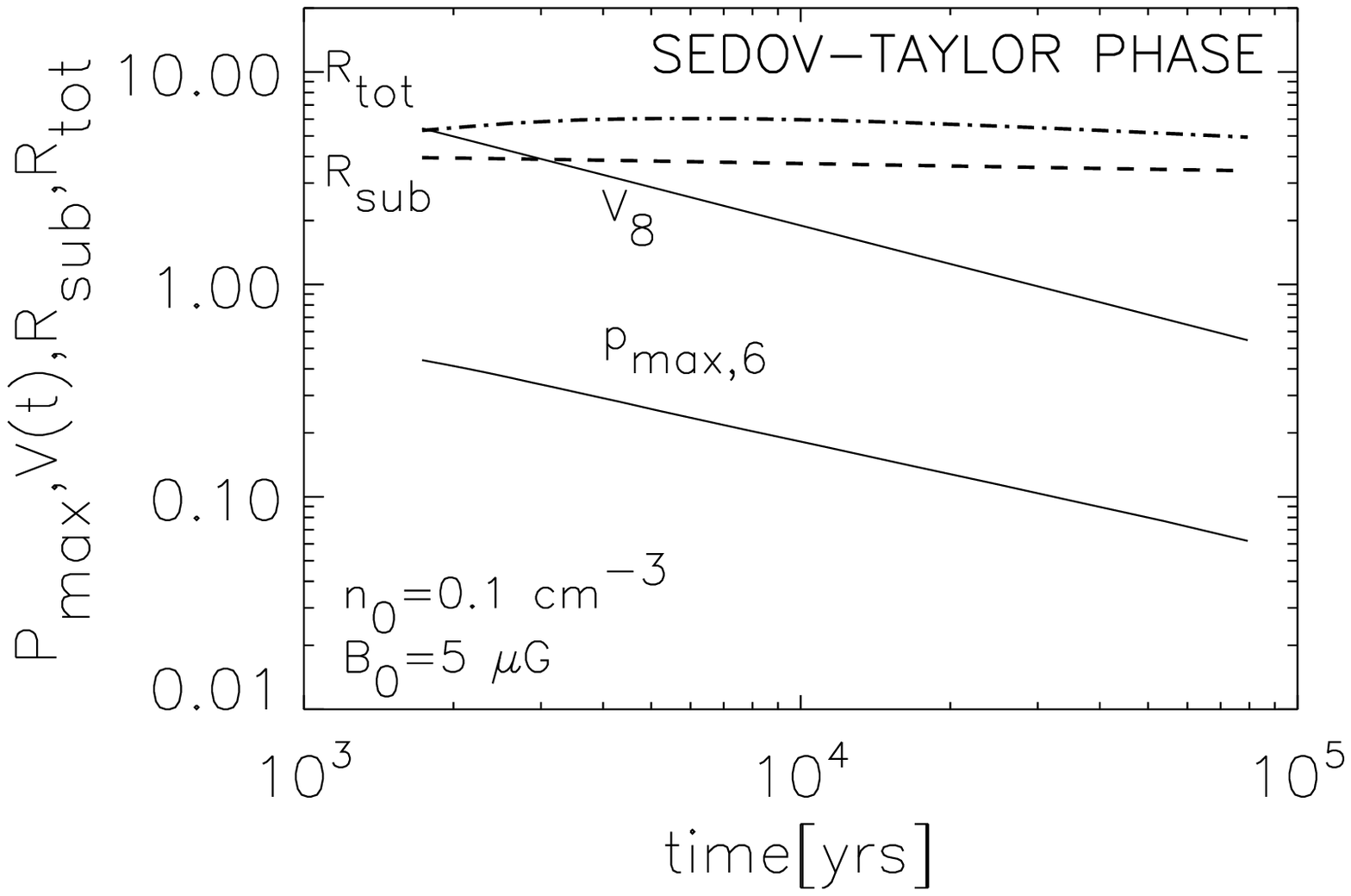}
\includegraphics{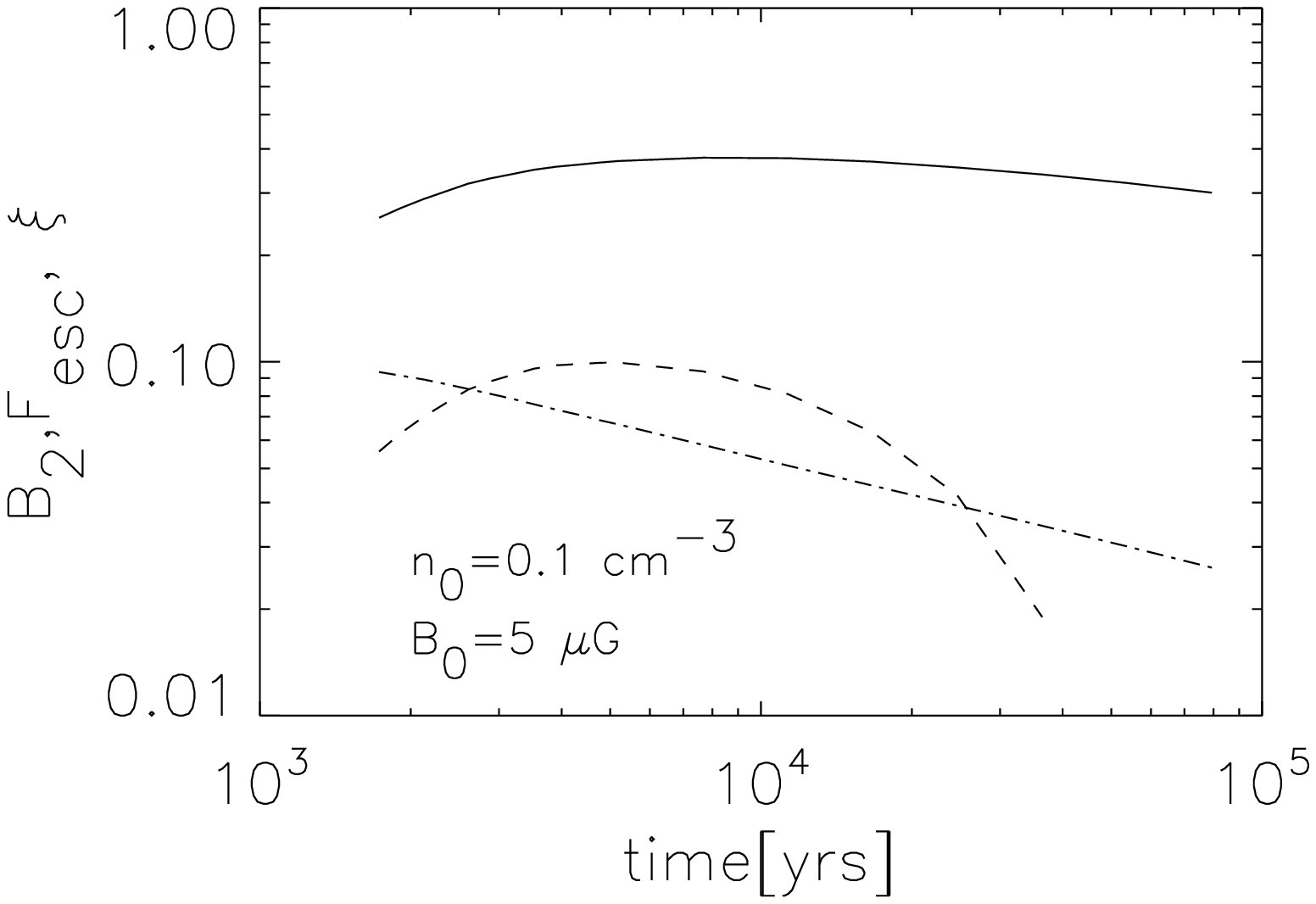}
}
\caption{Same as Fig.~\ref{fig:B1n01} but for $B_0=5\ \mu G$ and
  $n_0=0.1\ {\rm 
cm}^{-3}$.
}
\label{fig:B5n01}
\end{figure}

\begin{figure}
\resizebox{\hsize}{!}{
\includegraphics{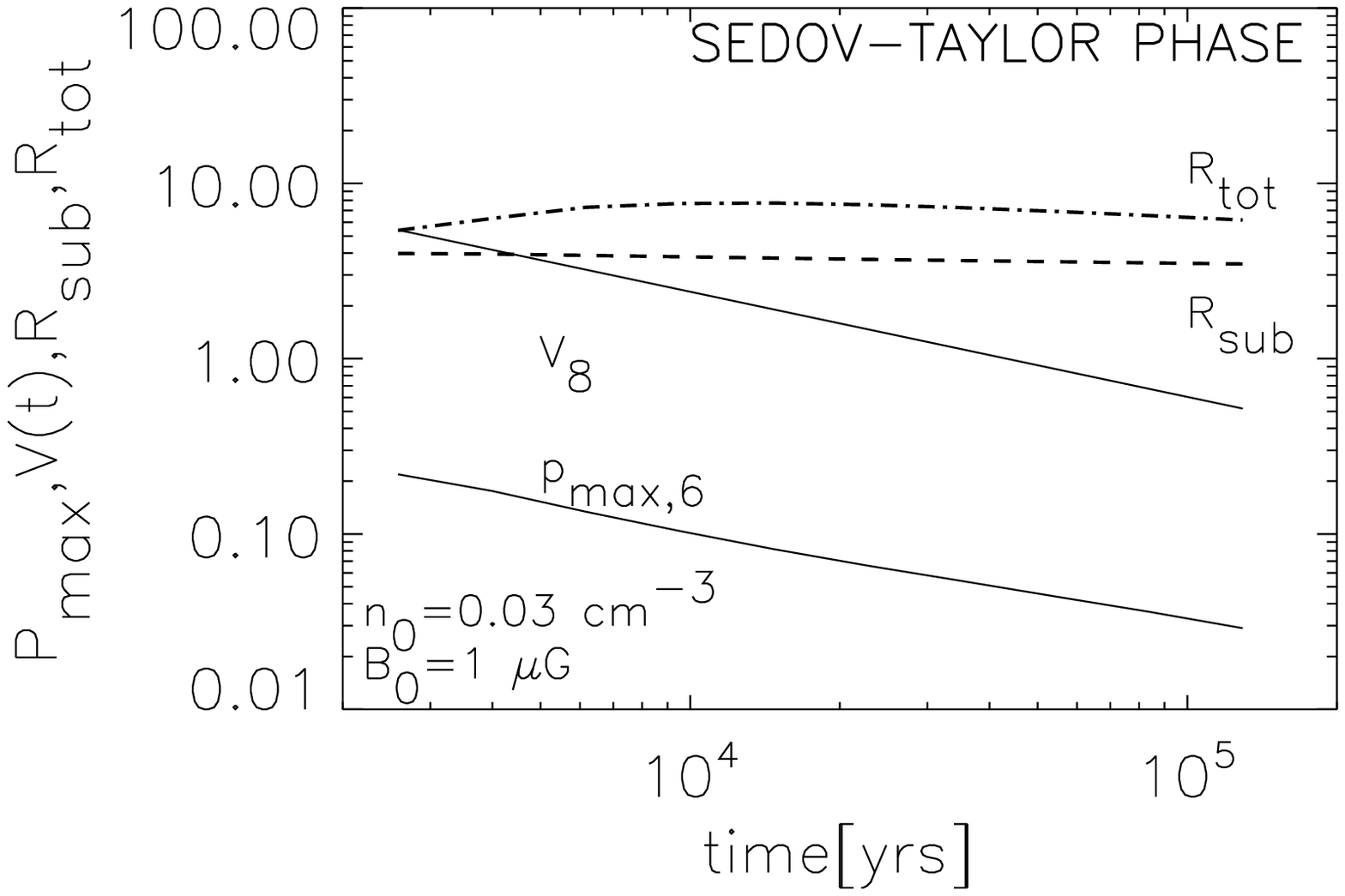}
\includegraphics{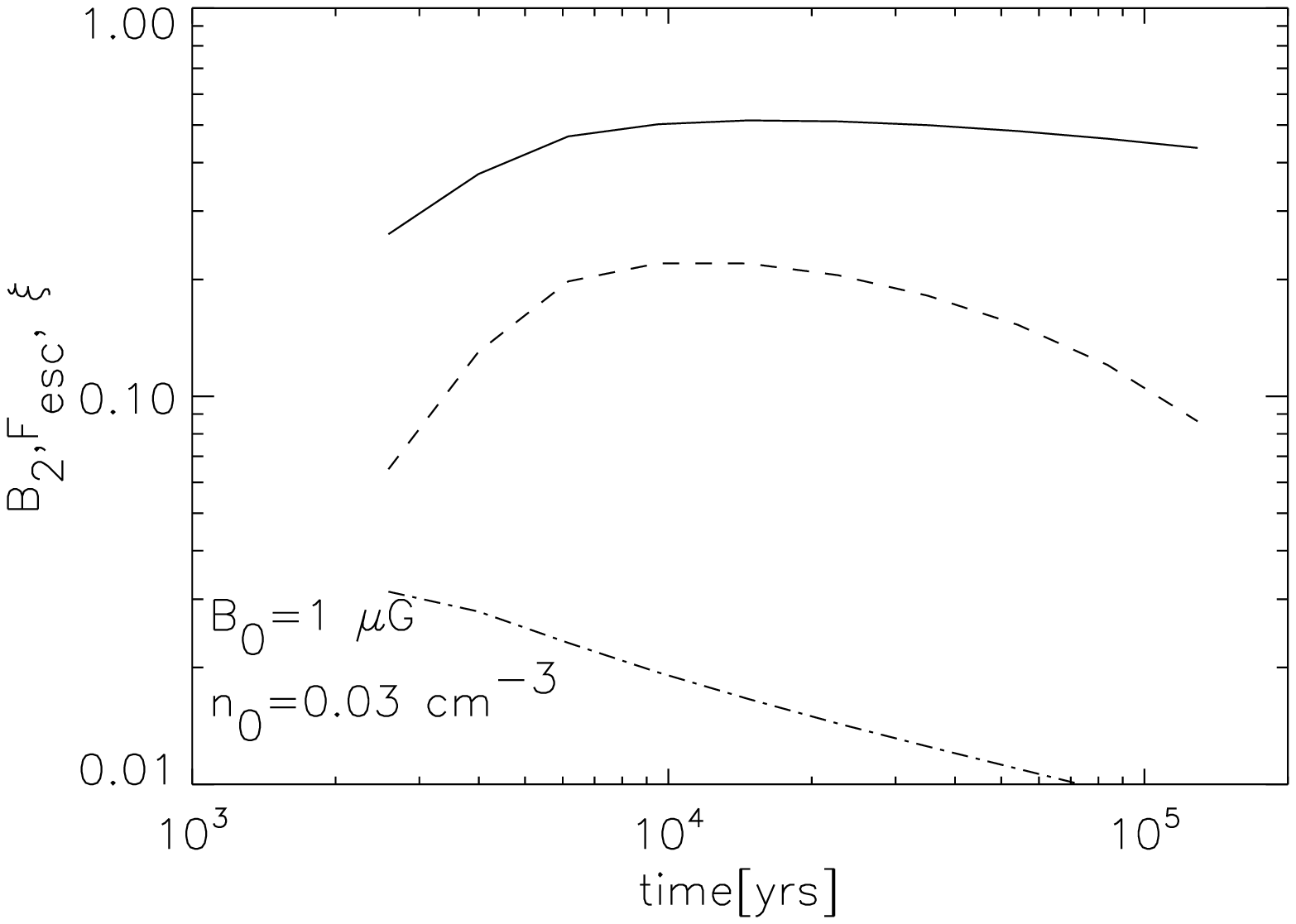}
}
\caption{Same as Fig.~\ref{fig:B1n01} and Fig.~\ref{fig:B5n01} but 
for $B_0=1\ \mu G$ and $n_0=0.03\ {\rm cm}^{-3}$.
}
\label{fig:B1n003}
\end{figure}

Our complete calculations, including the non-linear dynamical reaction of
the accelerated particles, the resonant amplification of magnetic field 
and the dynamical reaction of the field itself have been carried out as 
discussed by \cite{long}. The results are illustrated in 
Figs.~\ref{fig:B1n01}-\ref{fig:B1n003}. In the left panels we plot the maximum
momentum ($p_{max}$), the shock velocity ($V$) and the two compression factors
($R_{sub}$ and $R_{tot}$) as functions of time. The right panels show the
acceleration efficiency and the escape flux normalized to $\rho_0 V_{sh}^2$
and $(1/2)\rho_0 V_{sh}^3$ respectively, and the strength of the downstream
magnetic field in units of $10^3\mu G$.
The maximum momentum and the shock speed are in units of $10^6 m_p c$ and $10^8
\rm cm\,s^{-1}$ respectively. The three figures refer to the following sets of
parameters: $n_0=0.1\rm cm^{-3}$, $B_0=1\mu G$ (Fig.~\ref{fig:B1n01}),
$n_0=0.1\rm cm^{-3}$, $B_0=5\mu G$ (Fig.~\ref{fig:B5n01}) and $n_0=0.03\rm 
cm^{-3}$, $B_0=1\mu G$ (Fig.~\ref{fig:B1n003}).

The maximum momentum is determined at each time by requiring that the diffusion
length in the upstream section equals $0.1 R_{sh}(t)$. The
quantities $p_{max}(t)$, $R_{sub}(t)$ and $R_{tot}(t)$ are all outputs of the
non-linear calculations at the time $t$. The first point in time in
all figures corresponds to the beginning of the Sedov phase. The time
at which the Sedov-Taylor expansion begins was determined assuming
$E_{51}=1$ and that the mass of the ejecta is $M_{ej}=5 M_{\sun}$. The other 
relevant parameters are the temperature of the ISM in which the SNR is
expanding, for which we assumed $T_0=10^4\ K$, and the momentum threshold
for particles to be injected into the accelerator, which was chosen to
be $p_{inj}=\psi_{inj} \sqrt{2 k_B T_2 m_p}$, with $\psi_{inj}=3.8$ and 
$T_2$ the temperature downstream of the shock (see \cite{long} for details).

Some general comments are in order: one may notice that the total compression
factors obtained in our calculations are always lower than $\sim 10$. This is
uniquely due to the dynamical reaction of the amplified magnetic field. As
shown by \cite{apjlett} the effect of the amplified field on the plasma 
compressibility is relevant whenever the magnetic pressure becomes comparable 
with the thermal pressure of the background plasma upstream. The consequent
decrease in the compression ratios allows us to be consistent with the 
values that have been inferred from observations of a few SNRs 
(\cite{warren}). 

The highest momentum of accelerated particles, as expected, is reached at the
beginning of the Sedov-Taylor phase and is of order $\sim 10^6$ GeV (about the
knee). This should be considered as a lower limit to the maximum
energy reached 
at that time, since we have decided not to include the non-resonant channel of
magnetic field amplification, which is very efficient when the shock velocity 
is large. During the following expansion, the time-dependence of $p_{max}$
is reasonably well approximated by $p_{max} \propto t^{-1/2}$, in agreement 
with Eq.~\ref{eq:emax1t}, since $\xi_c(t)$ is roughly constant (see solid 
curve on the right panels of 
Figs.~\ref{fig:B1n01}-\ref{fig:B5n01}-\ref{fig:B1n003} and discussion below).

A crucial ingredient in calculating the maximum energy at a given time is the
strength of the magnetic field. The magnetic field intensity in the
downstream  
plasma is plotted in the right panels (dot-dashed line) for the three cases 
considered here. The typical values are between $\sim {\rm a\,few -}10\mu G$ 
at late times and $\sim 30-100\mu G$ at the beginning of the Sedov
phase. After the first few thousand years, the
dependence on time is not far from $\delta B_2 \propto t^{-3/10}$,
as would result from Eq.~\ref{eq:delB1t}, using the additional information
on the approximate constancy of $\xi_c(t)$ and $R_{sub}$ (dashed line in
the left panels of Figs.~\ref{fig:B1n01} to \ref{fig:B1n003}). Despite
these resemblances, the situation to which the plots refer is more
complicated than that described by Eq.~\ref{eq:delB1t}, where a number
of effects have been neglected, first among these the presence of a
precursor which evolves with time ($R_{tot}$ is changing as can be
seen from the dot-dashed curve on the left of Figs.~\ref{fig:B1n01} to
\ref{fig:B1n003}) and the time varying adiabatic compression it
entails.

The right panels also show the acceleration efficiency (solid line) and
normalized escape flux (dashed lines). One should notice that even when
the acceleration efficiency is very high, of order $\sim 50-60\%$, the 
escape energy flux never exceeds $\sim 30\%$. As discussed above, this 
latter quantity should be the one that is more directly related to the 
cosmic ray energetics in the Galaxy, at least at the highest energies, 
while the former is more relevant for the generation of secondary 
radiation due to cosmic ray interactions in the acceleration region. 

The acceleration efficiency and the normalized escape flux initially 
increase with time during the Sedov-Taylor expansion phase. This
behaviour is related to an analogous trend of the shock modification, 
as can be clearly seen from the time-dependence of $R_{tot}$
(dash-dotted curve in the left panel of 
Figs.~\ref{fig:B1n01}-\ref{fig:B1n003}). In fact, at the beginning of
the Sedov phase, the amplified magnetic field is at a maximum and its 
dynamical reaction on the shock is so strong that the acceleration 
efficiency is reduced. As soon as the magnetic field strength starts 
decreasing, the shock modification increases, and $\xi_c$ and
$F_{esc}$ with it. Notice, however,  that 
this does not mean that the actual cosmic ray pressure and escape flux
increase, because $\xi_c$ and $F_{esc}$ are normalized to 
$\rho_0 V_{sh}^2(t)$ and $\rho_0 V_{sh}^3(t)/2$ respectively, 
and both decrease with time rather quickly.

At later times, both $\xi_c$ and $F_{esc}$ start decreasing, with the
latter showing a more rapid decline than the former. This is due to the 
fact that the shock is slowing down and progressively becoming
unmodified: the maximum momentum is decreasing and the spectrum of 
accelerated particles is steepening. Recalling again that the plots
show normalized quantities, one gathers that the decline with time of 
cosmic ray pressure and escaping energy flux is quite dramatic in this
phase.

\section{Conclusions}
\label{sec:concl}

The escape of accelerated particles from the shock region towards
upstream appears in basically all approaches to non-linear particle
acceleration at shock fronts. In kinetic approaches that solve the
relevant equations assuming quasi-stationarity at any given time the
escape flux appears either as a corrective term in the equation of
energy conservation between downstream and upstream infinity, or as a
result of imposing as a boundary condition that the distribution
function of accelerated particles vanishes at a finite distance $x_0$
upstream of the shock. In this second case the escape flux is
described by the non-vanishing spatial derivative of the distribution
function at $x_0$. Such a quasi-stationary description of the
acceleration process is meaningful only whenever the shock velocity
decreases in time and the magnetization at the shock, as due to
streaming of cosmic rays, also decreases in time. These two conditions
lead to escape of the particles with momentum close to the highest
achievable at any given time. This would be the case during the
Sedov-Taylor phase of the expansion of a supernova shock. These
approaches are therefore unfit to describe the phase of free
expansion, where the maximum momentum increases with time and no
escape towards upstream takes place. 

Fully time-dependent, numerical calculations apply to arbitrary
situations and should be able to describe the increase of momentum
during the free expansion as well as the escape of particles during
the Sedov phase. However, in most cases these calculations do not
include the time dependence of the magnetic field at the shock. This
leads to an increases of the maximum momentum even during the Sedov
phase. Probably for this reason there is no detailed discussion in the
literature of the escape flux from a SNR in the context of 
time dependent calculations. The appearance of the escape flux in
numerical time dependent approaches would depend on the boundary
conditions adopted in the specific case. If the distribution function
is assumed to vanish at upstream infinity, the escaping particles
should appear as a concentration of particles with large momenta
(close to $p_{max}(t)$) at large distances from the shock: since the
acceleration box extends to infinity, there is no real
escape. On the other hand, if the boundary condition $f(x_0)=0$ is
imposed at a finite distance from the shock, the escape flux (as a
function of time) should appear automatically as a consequence of a
non-vanishing $D\partial f(x_0)/\partial x$. It would be interesting to
have confirmation of these expectations by running time dependent
calculations with decreasing magnetization.

The existence of the escape flux is not simply a mathematical
nuisance: as discussed by \cite{ptuzira05} it is the very reason why
supernova remnants can potentially be the sources of cosmic rays up to
the knee. If the particles we observe as cosmic rays were the ones
advected towards downstream of the shock, the adiabatic losses
suffered during the expansion of the shell would drive these particles
towards lower energies. It is therefore of the highest importance to
go beyond the test particle calculation of \cite{ptuzira05} and achieve
an understanding of the escape flux in the context of the fully
non-linear theory of diffusive shock acceleration.
 
In this paper we made a first step in this direction, by calculating
the temporal evolution of the escape flux during the Sedov-Taylor
phase of the expansion of a supernova in a medium with constant
density. Our calculation included a recipe to describe the cosmic ray
induced magnetic field, the dynamical reaction of the accelerated
particles and the field itself. In addition to this we also discussed
in some detail the limitations of the conservation equations as they
are usually written, in that they do not allow to describe the
magnetic field perturbations when they are not Alfv\'en waves.  

We showed that the escape flux may involve between few 
and 10-30 \% of the shock ram pressure, while the particle acceleration
efficiency at the same time reaches 50-60 \%. This means that the
energetics inferred from observations of secondary radiation from a
remnant, for instance in the form of gamma rays, may not be the
relevant one for the origin of cosmic rays, since it is not related in
a trivial way to the energy flux escaping the accelerator.

The maximum energy up to which particles may get accelerated is reached at the
beginning of the Sedov phase and is of order $10^{15}$ eV if only resonant
amplification of the field is included. $E_{max}$ might be somewhat
larger for some SNRs that at the very beginning of the Sedov-Taylor
phase may experience the effect of non-resonant streaming instability
\cite[]{bell04}. This mechanism provides extremely efficient field
amplification as long as the shock velocity is high, and hence is
expected to play a very important role in the early 
Sedov-Taylor phase (and possibly during the free expansion phase). At present,
non-resonant modes cannot be formally accounted for in the conservation
equations, since a detailed description of the energy transfer between 
particles and waves is not available yet. We carried out all calculations 
in the simpler case of Alfv\'en waves interacting with particles in a 
resonant way. 

The amplification of the magnetic field, by either resonant or non-resonant
streaming instability, has profound implications on the escape flux of
particles towards upstream of a shock, and therefore on the spectrum of cosmic
rays we observe at the Earth. The most obvious consequence of the
magnetic field amplification is that of allowing for higher values of the
maximum energy of accelerated particles, as shown by our Eqs.~\ref{eq:emax1t}
and \ref{eq:emax2t}. However large magnetic fields exert a dynamical reaction
on the plasma leading to a reduction of the compression in the precursor. This
happens whenever the magnetic pressure exceeds the pressure of the background
gas \cite[]{apjlett}. As a result, the concavity of the spectrum of
accelerated 
particles \cite[]{long} is reduced and at the same time the escape flux at $p\sim
p_{max}$ decreases. It follows that larger field implies larger $p_{max}$ but not
necessarily larger escape flux, as shown in Figs.~\ref{fig:B1n003},
\ref{fig:B1n01} and \ref{fig:B5n01} (see the behaviour of the curves at early
times). 

The escape of accelerated particles from a cosmic ray modified shock has
profound implications for the origin of cosmic rays, which will be
discussed in detail in a forthcoming paper. Here we want to emphasize
some general points: 

1) the escape from upstream is the natural solution to the well known problem
of explaining how the highest energy particles (say, close to the knee energy)
could escape the system without suffering substantial adiabatic energy losses;

2) the magnetic field amplification is expected to switch from mainly
non-resonant to mainly resonant at the beginning of the Sedov-Taylor phase. It
can be easily understood that this may lead to peculiar changes in the
spectrum of cosmic rays detected at the Earth, reflecting this transition;

3) the flux of escaping particles, once integrated in time during the SNR
evolution may be very different from the concave instantaneous spectrum which
can potentially be observed in a SNR, for instance by looking at its
multifrequency emission. This point is certainly relevant for the
purpose of addressing the commonly raised point of how a concave
spectrum  of accelerated particles can reflect in an almost perfect
power law over many orders of magnitude;

4) there is a further complication of all the picture, due to the acceleration
of nuclei at energies that may be expected to scale as the charge of the
nucleus (in the case of Bohm diffusion). Any calculation of the flux of single
chemical species observed at the Earth must take these complex effects into
account. 

\section*{Acknowledgments}
We are grateful to G. Cassam-Chenai and S. Gabici for reading
an intermediate version of the present manuscript. We are also
grateful to the anonymous referee for his/her precious comments. 
This work was partially supported by PRIN-2006, by ASI through
contract ASI-INAF I/088/06/0 and (for PB) by the US DOE and by NASA
grant NAG5-10842. Fermilab is operated by Fermi Research Alliance, LLC
under Contract No. DE-AC02-07CH11359 with the United States DOE.

\bibliographystyle{mn2e}
\bibliography{escapelast}

\end{document}